\documentclass[reprint,superscriptaddress,amsmath,amssymb]{revtex4-2}
\usepackage{graphicx}% Include figure files
\usepackage{dcolumn}% Align table columns on decimal point
\usepackage{bm}% bold math
\usepackage{braket}
\usepackage{appendix}
\usepackage{subfigure}
\usepackage{hyperref}
\usepackage{color}
\usepackage{xcolor}
\usepackage{amsmath}
\usepackage{nccmath}
\usepackage[english]{babel}
\usepackage{svg}
\usepackage{lineno}
\begin{document}

	\title{Nanotube spin defects for omnidirectional magnetic field sensing}% Force line breaks with \\
	
	\author{Xingyu Gao}
	\thanks{These authors contributed equally to this work.}
	\affiliation{		Department of Physics and Astronomy, Purdue University, West Lafayette, Indiana 47907, USA}

  \author{Sumukh Vaidya}
  \thanks{These authors contributed equally to this work.}
\affiliation{	Department of Physics and Astronomy, Purdue University, West Lafayette, Indiana 47907, USA}

	\author{Saakshi Dikshit}%
\affiliation{	Elmore Family School of Electrical and Computer Engineering, Purdue University, West Lafayette, Indiana 47907, USA}%
	
	\author{Peng Ju}
	\affiliation{		Department of Physics and Astronomy, Purdue University, West Lafayette, Indiana 47907, USA}

	\author{Kunhong Shen}%
	\affiliation{		Department of Physics and Astronomy, Purdue University, West Lafayette, Indiana 47907, USA}

    \author{Yuanbin Jin}%
	\affiliation{		Department of Physics and Astronomy, Purdue University, West Lafayette, Indiana 47907, USA}

    \author{Shixiong Zhang}%
\affiliation{	Department of Physics, Indiana University, Bloomington, Indiana 47405, USA}
\affiliation{	Quantum Science and Engineering Center, Indiana University, Bloomington, IN 47405, USA}

	\author{Tongcang Li}%
	\email{tcli@purdue.edu}
	\affiliation{		Department of Physics and Astronomy, Purdue University, West Lafayette, Indiana 47907, USA}
	\affiliation{		Elmore Family School of Electrical and Computer Engineering, Purdue University, West Lafayette, Indiana 47907, USA}
	\affiliation{		Purdue Quantum Science and Engineering Institute, Purdue University, West Lafayette, Indiana 47907, USA}
	\affiliation{		Birck Nanotechnology Center, Purdue University, West Lafayette, Indiana 47907, USA}
	\date{\today}% It is always \today, today,

\begin{abstract}
%	\normalsize \bf
Optically addressable spin defects in three-dimensional (3D) crystals and two-dimensional (2D) van der Waals (vdW) materials are revolutionizing nanoscale quantum sensing. Spin defects in one-dimensional (1D) vdW nanotubes will provide unique opportunities due to their small sizes in two dimensions and absence of dangling bonds on side walls. However, optically detected magnetic resonance of localized spin defects in a nanotube has not been observed. Here, we report the observation of single spin color centers in boron nitride nanotubes (BNNTs) at room temperature. Our findings suggest that these BNNT spin defects possess a spin $S=1/2$ ground state without an intrinsic quantization axis, leading to orientation-independent magnetic field sensing. We harness this unique feature to observe anisotropic magnetization of a 2D magnet in magnetic fields along orthogonal directions, a challenge for conventional spin $S=1$ defects such as diamond nitrogen-vacancy centers. Additionally, we develop a method to deterministically transfer a BNNT onto a cantilever and use it to demonstrate scanning probe magnetometry. Further refinement of our approach will enable atomic scale quantum sensing of magnetic fields in any direction. 
\end{abstract}

\maketitle

%\textbf {Keywords:} {spin defects, hexagonal boron nitride, quantum sensing, optically detected magnetic resonance, paramagnetic ions}

%\section{Main}

\begin{figure*}[tph]
	\centering
	\includegraphics[width=0.98\textwidth]{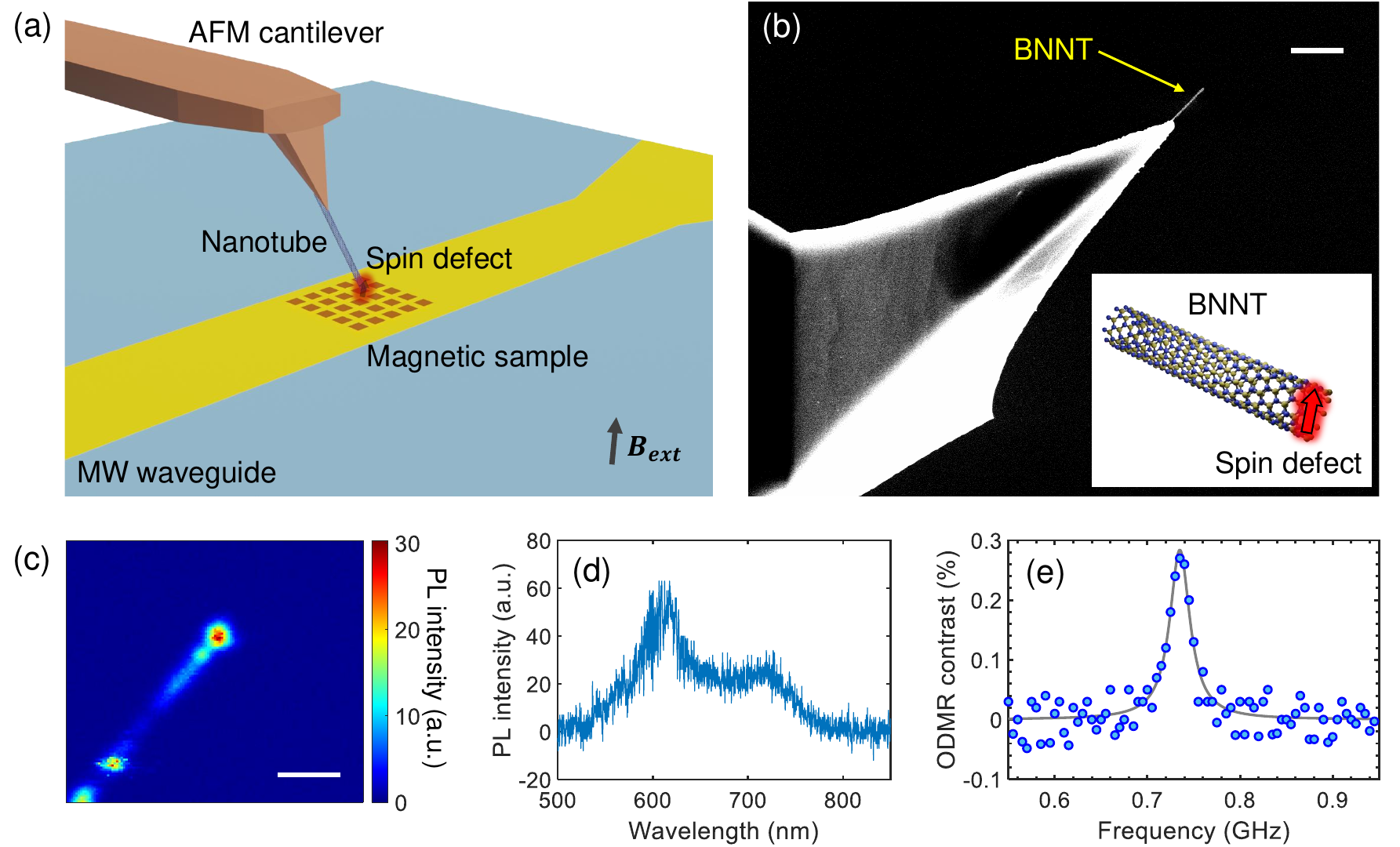}
	\caption{ \textbf{Spin defect in a nanotube affixed to an AFM cantilever.} (a) Conceptual schematic of scanning-probe magnetometry with an optically-active spin defect in a nanotube attached to an AFM cantilever. The spin defect is excited by a laser beam and by a microwave (MW). The resulting photoluminescence (PL) is collected by a confocal system. As the sample on the MW waveguide is scanned, ODMR measurements can be conducted to create a 2D map of the stray magnetic field generated by the sample.  (b) An SEM image of a single BNNT attached to the tip of an AFM cantilever. Scale bar: 2 $\mu$m. Inset: An illustration of a spin defect  at the end of a BNNT. (c) A confocal microscopy image of a BNNT mounted on a cantilever. The bright spot at the end of the nanotube signifying an optically-active spin defect. Scale bar: 2 $\mu$m. a.u.: arbitrary unit. (d) A PL spectrum of the spin defect shown in subfigure (c). A 532 nm laser is used to excite the defect. (e) An example ODMR spectrum taken with the spin defect on the cantilever. An external magnetic field of 26 mT is applied in the measurement. The MW waveguide is placed approximately 100 $\mu$m away from the BNNT probe.  } \label{fig1:AFM}
\end{figure*}

Optically addressable spin defects in solids \cite{wolfowicz2021quantum} are outstanding platforms for quantum sensing \cite{schirhagl2014nitrogen,shi2015single} and other quantum information applications \cite{stas2022robust,weber2010quantum}.  A prominent example of spin defects in three-dimensional (3D) crystals is the diamond nitrogen-vacancy (NV) center \cite{doherty2013nitrogen}. These diamond NV centers enable nanoscale measurements of magnetic field, temperature and other external stimuli, thus opening up a broad range of applications in condensed matter physics, biology, and material science \cite{balasubramanian2008nanoscale,thiel2019probing,song2021direct,aslam2023quantum}. However, the fabrication of diamond-based devices remains challenging and costly, and the properties of NV centers degrade when near diamond surfaces. Recently, the discovery of optically-active spin defects in hexagonal boron nitride (hBN) \cite{gottscholl2020initialization,mendelson2021identifying,chejanovsky2021single}, a two-dimensional (2D) van der Waals (vdW) material, offers new opportunities in the field of   quantum sensing \cite{gottscholl2021spin,healey2022quantum,huang2022wide,gao2023quantum,robertson2023detection,vaidya2023quantum}. The 2D nature of hBN allows  the presence of stable  spin defects in atomically-thin flakes \cite{durand2023optically}, enabling them to be in close proximity to the target sample, thereby enhancing the signal. Most studies on hBN spin defects have focused on ensembles of negatively charged boron vacancies (V$_B^-$)  \cite{gottscholl2021room,gao2022nuclear,haykal2022decoherence,mathur2022excited,gong2023coherent,rizzato2023extending,ru2023robust}. The V$_B^-$ ensembles have been used to image 2D magnets  with wide-field microscopy \cite{healey2022quantum,huang2022wide} but are limited by the optical diffraction limit.
Furthermore, both diamond NV centers and hBN V$_B^-$ defects are spin $S=1$ defects with intrinsic spin axes, which makes them insensitive to off-axis magnetic fields \cite{foglszinger2022tr12}. Electron spin qubits in quantum dots, which are zero-dimensional (0D) materials, have been used for quantum information processing at cryogenic temperatures \cite{press2008complete}. But their spin lifetime is very short at room temperature, which limits their applications in quantum sensing.

One-dimensional (1D) vdW nanotubes with optically-active spin defects will provide unique opportunities for quantum sensing, particularly when configured into sharp tips for scanning probe  magnetometry (Fig. \ref{fig1:AFM}(a)). Because vdW nanotubes are free of dangling bonds on their side walls, they can remain robust even at extremely small diameters. This feature has led to the use of carbon nanotube tips in atomic force microscopy (AFM) for high-resolution imaging of complex surface topographies \cite{wilson2009carbon}. In addition, spin defects in a nanotube can be coupled to the nanotube's mechanical vibration for quantum information transduction \cite{rabl2010quantum,palyi2012spin,li2012spin}. Spin qubits in carbon nanotubes have been observed by electronic transport \cite{mason2004local} and conventional electron paramagnetic resonance \cite{chen2023long} at low temperatures. However, optically detected magnetic resonance (ODMR)  of localized spin defects in nanotubes, crucial for quantum sensing, remains elusive.

In this article, we report the observation of single optically-active spin defects in boron nitride nanotubes (BNNTs) at room temperature. BNNTs  are structurally similar to carbon nanotubes but have a large band-gap ($\sim$ 6 eV) \cite{golberg2010boron,xu2022advances}, which enables them to host quantum emitters \cite{chejanovsky2017quantum,ahn2018stable}. Our ODMR measurements reveal negligible zero-field splitting (ZFS) in the ground state of these defects, suggesting the lack of spin quantization axis in the absence of an external magnetic field. Rabi oscillation measurements further confirm that these BNNT spin defects have a spin $S= 1/2$ ground state. Remarkably, we find that the ODMR resonant frequency and contrast are independent of the magnetic field orientation, offering nearly constant omnidirectional sensitivity for magnetic field detection. To show the capability of BNNT spin defects in omnidirectional magnetic field sensing, we perform measurements of anisotropic magnetization of a representative 2D magnet. Moreover, we develop an efficient technique to deterministically transfer a single BNNT, which contains spin defects,  onto the tip of an AFM cantilever. Utilizing a BNNT-attached cantilever as a probe for scanning-probe ODMR measurements, we measure the stray magnetic field distribution near nickel patterns. These findings highlight the great potential of BNNT spin defects for high resolution sensing applications. 

{\bf Observation of spin defects in BNNTs}

A conceptual schematic of scanning probe magnetometry with a spin defect in a nanotube is shown in Figure \ref{fig1:AFM}(a). A BNNT, containing a spin defect at its  end, is attached to the tip of an AFM cantilever. The BNNT spin defect is excited using a 532-nm laser, and its spin-dependent photoluminescence (PL) is monitored as the underlying sample is scanned.  A gold microstripe antenna underneath the sample delivers a microwave (MW) to drive electron spin transitions and enable the acquisition of spectra. A nearby permanent magnet provides an external magnetic field, $B_{ext}$, both to separate the spin sublevels of the BNNT spin defects and to magnetize the sample. 

\begin{figure*}[tbhp]
	\centering
	\includegraphics[width=0.75\textwidth]{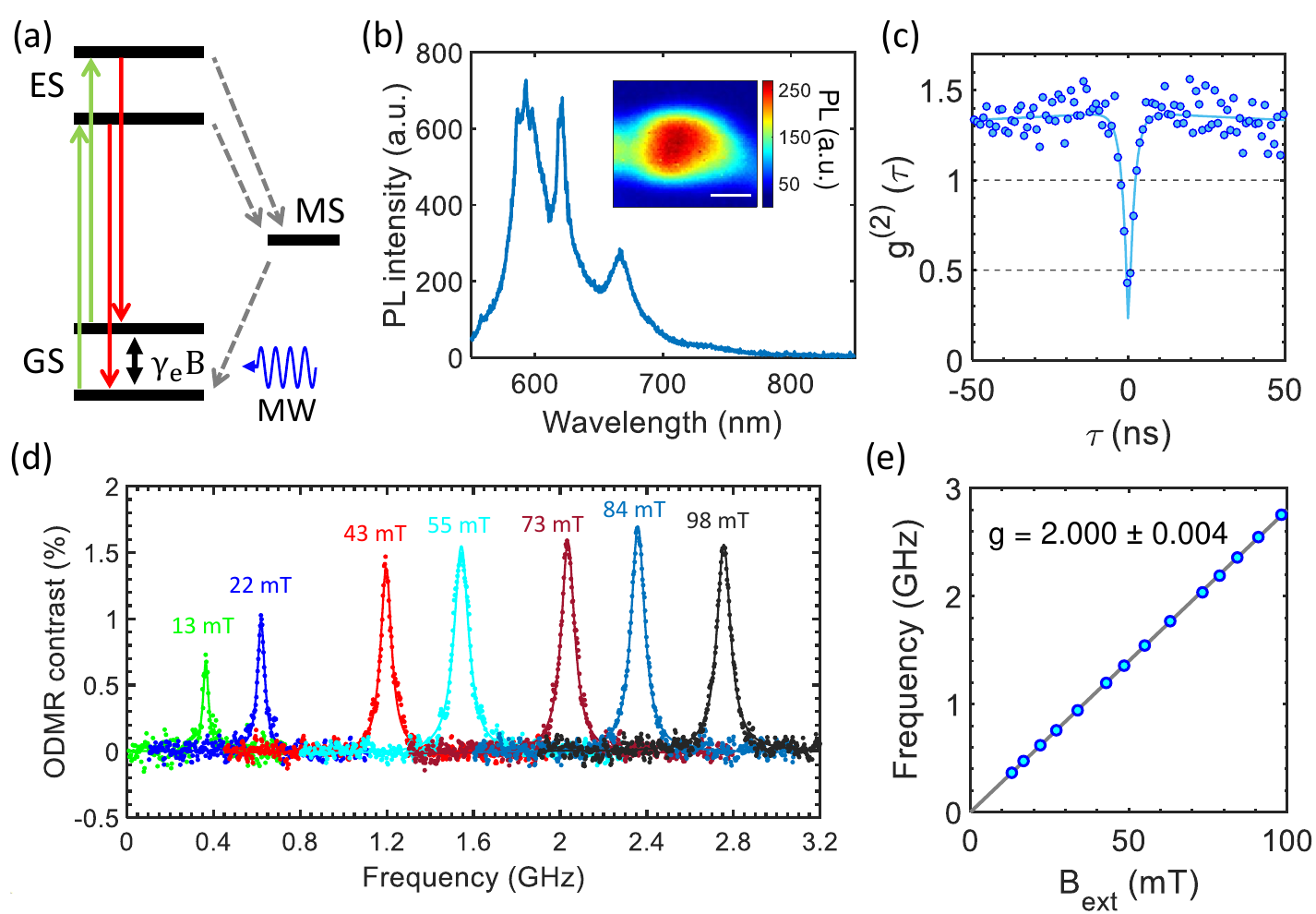}
	\caption{ \textbf{A single spin defect in BNNT.} (a) A possible energy level diagram depicting the electron spin dynamics under laser and microwave excitation. The spin defect contains a spin doublet ground state (GS), doublet excited state (ES) and a metastable state (MS). The greens arrows are the spin conservative excitation under 532 nm laser pumping. The red arrows are spontaneous emission from ES to GS. Gray dashed lines are non-radiative decays, known as intersystem crossing. An external magnetic field can cause a splitting of $\gamma_eB$ in the ground state via the Zeeman effect. (b) An optical spectrum of an isolated spin defect in a BNNT. Inset: an confocal PL map of this isolated defect. The scale bar is 500 nm. (c) Second-order intensity-correlation measurement for the defect at 400 $\mu$W laser excitation. The solid line is a theoretical fit. (d) ODMR spectra taken in different magnetic fields. (e) The fitted resonance frequency from (d) as a function of the magnetic field $B_{ext}$. The solid line is a linear fitting of the data, showing a slope of 2.000$\pm$0.004.  } \label{fig2:BNNT}
\end{figure*}

In our experiments, we observe that BNNTs, with an average diameter of 50 nm (Supplemental Fig. S1) \cite{ahn2018stable,starko2019high}, host stable optically addressable spin defects.  To  construct a BNNT scanning probe, we first place BNNTs on a flat cantilever (Supplemental Fig. S2). We then use an AFM cantilever with an extended tip to pick up a nanotube with spin defects (Supplemental Fig. S2, S3).  Figure \ref{fig1:AFM}(b) shows a scanning electron microscope (SEM) image of a BNNT attached to an AFM cantilever. The nanotube sticks out by several micrometers, facilitating the  collection of photons emitted by the spin defects. A PL map (Figure \ref{fig1:AFM}(c)), taken by our home-built confocal microscope, shows a bright emitter located at the nanotube's tip. This emitter produce room temperature PL in the visible spectrum (Figure \ref{fig1:AFM}(d)). More importantly, this PL is spin-dependent. During the continuous-wave (CW) ODMR measurement, we record PL count rates under a 532-nm laser excitation as a function of the applied MW frequency. By modulating the MW with a 1 kHz square wave, we determine the ODMR contrast ($\Delta$PL/PL) at each MW frequency by gauging the normalized change in PL count rates between the situations of MW being on and off. Figure \ref{fig1:AFM}(e) presents a representative ODMR spectrum in a 26-mT magnetic field. The spin defect shows a resonant peak at 735 MHz with a 26 MHz linewidth. A positive ODMR contrast indicates an increase of PL intensity when the MW is applied at the spin resonance frequency \cite{stern2022room}. 

\begin{figure*}[htbp]
	\centering
	\includegraphics[width=0.7\textwidth]{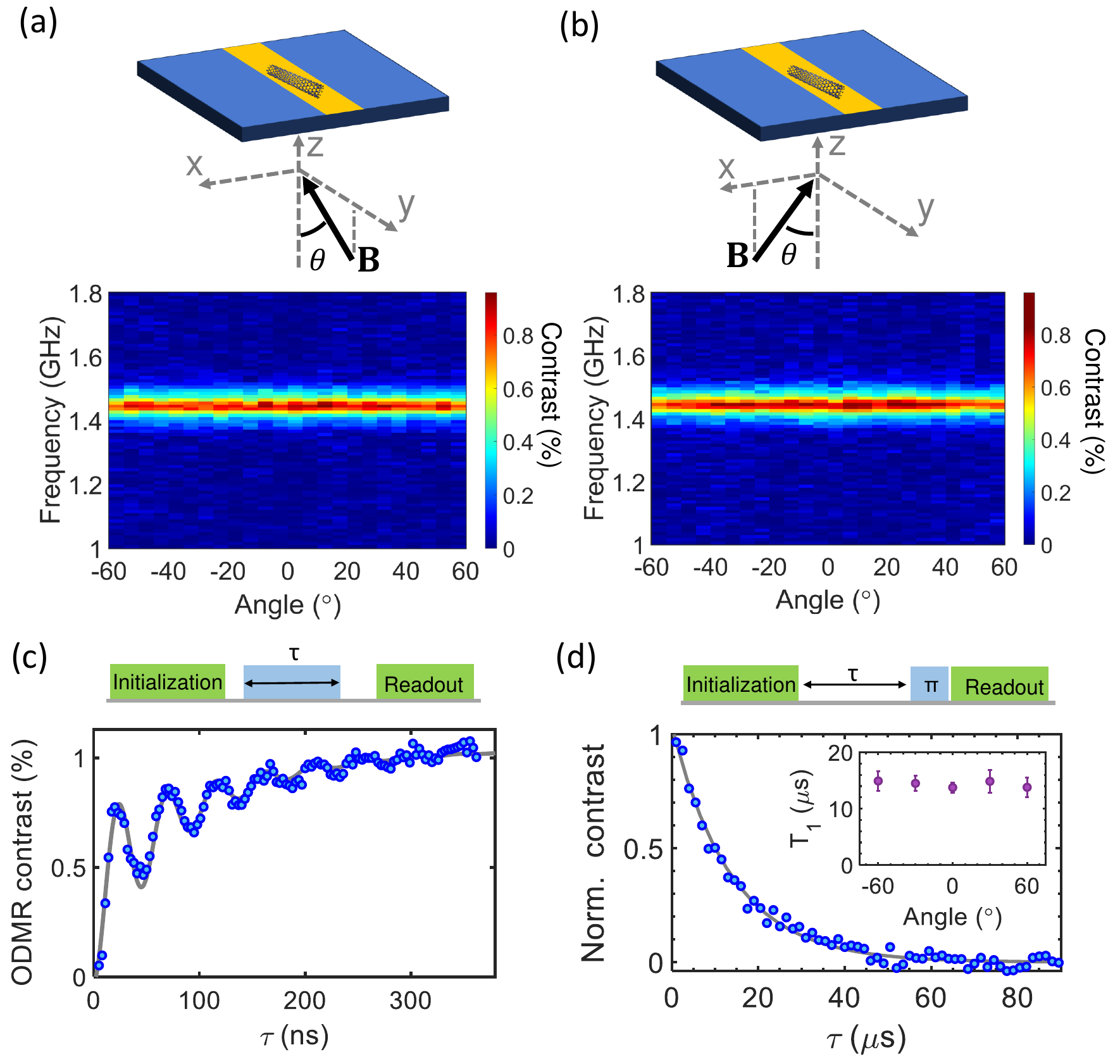}
	\caption{ \textbf{ODMR measurements at different orientations.} (a)-(b) Top panels: Illustrations of the magnetic field rotated around two  orthogonal axes.  Bottom panels: ODMR spectra as functions of the orientation of external magnetic fields. (c) Coherent control of the spin defect in BNNTs. The pulse sequence is shown in the top diagram. (d) $T_1$ relaxation measurement. The pulse sequence is shown in the top panel.  Inset: $T_1$ relaxation times for different magnetic field orientations.} \label{fig3:magnetic}
\end{figure*}

To gain more insight into the spin properties of BNNT spin defects, we characterize BNNT spin defects placed on a gold stripline MW waveguide at room temperature. Figure \ref{fig2:BNNT}(b) shows the PL spectrum of an isolated bright spot in a BNNT under 532-nm laser illumination. The measured second-order correlation, $g^{(2)}(\tau=0)$ $<$ 0.5, suggests that the photons primarily originate from a single defect (Figure \ref{fig2:BNNT}(c)). From the $g^{(2)}(\tau)$ measurement, the lifetime of the excited state is estimated to be $1.8 \pm 0.3$ ns. To pinpoint the ground state spin energy levels, we perform CW ODMR measurements across various out-of-plane magnetic fields ranging from 13 mT to 98 mT. As depicted in Figure \ref{fig2:BNNT}(d),  the spin defect exhibits a single spin resonance peak for each external magnetic field $B_{ext}$. The ODMR contrast first increases with increasing magnetic fields when the field is relatively weak ($B_{ext}<$ 40 mT), and then stabilizes at stronger fields ($B_{ext}>$ 40 mT).  If we consider this as a spin $S=1$ defect, its ground state spin Hamiltonian can be written as: $ H_{S=1} = D S_z^2 + g\mu_B\textbf{B}\cdot\textbf{S}$,
where $D$ is the ZFS parameter, $g$ is the Land\'e factor, and $\mu_B$ is the Bohr magneton. $\textbf{S}$ and $S_z$ are the spin operators. Linear fitting of the measured ODMR spin resonance frequency $\nu$ versus $B_{ext}$ yields a negligible ZFS (D$<$10 MHz) and a Land\'e factor of $g$=2.000$\pm$0.004 (Figure \ref{fig2:BNNT}(e)). This minimal ZFS points towards this defect possessing a spin $S=1/2$ ground state, which is further substantiated by comparing the Rabi frequency of this defect with that of the hBN $V_B^-$ defects  (Supplemental Fig. S5) \cite{scholten2023multi}.  An  energy level diagram of the BNNT spin defect is depicted in Figure \ref{fig2:BNNT}(a). For a spin $S=1/2$ defect, the ground state spin Hamiltonian is expressed as: 
\begin{equation}\label{eq2}
	H_{S=1/2} = g\mu_B\textbf{B}\cdot\textbf{S},
\end{equation}
which will be optimal for omnidirectional magnetic field sensing. 

{\bf Orientation-independent ODMR}

The negligible ZFS suggests that the observed BNNT spin defects have no intrinsic quantization axis. Consequently, the direction of their spin polarization is solely determined by the orientation of the external magnetic fields. As a result, the energy splitting (Eq. \ref{eq2}) only depends on the magnitude of a magnetic field, not its orientation. Given an appropriate MW drive (Supplemental Fig. S6), both the ODMR contrast and linewidth should remain unchanged regardless of the magnetic field orientation. This enables omnidirectional magnetic field sensing with nearly uniform sensitivity. This is a distinct advantage over anisotropic spin defects such as diamond NV centers and hBN $V_B^-$ defects. The latter ones have spin $S=1$  ground states with large ZFS (usually in the GHz range), which leads to a reduced ODMR contrast when the magnetic field does not align with their intrinsic quantization axis \cite{foglszinger2022tr12}. 

\begin{figure*}[htp]
	\centering
	\includegraphics[width=0.7\textwidth]{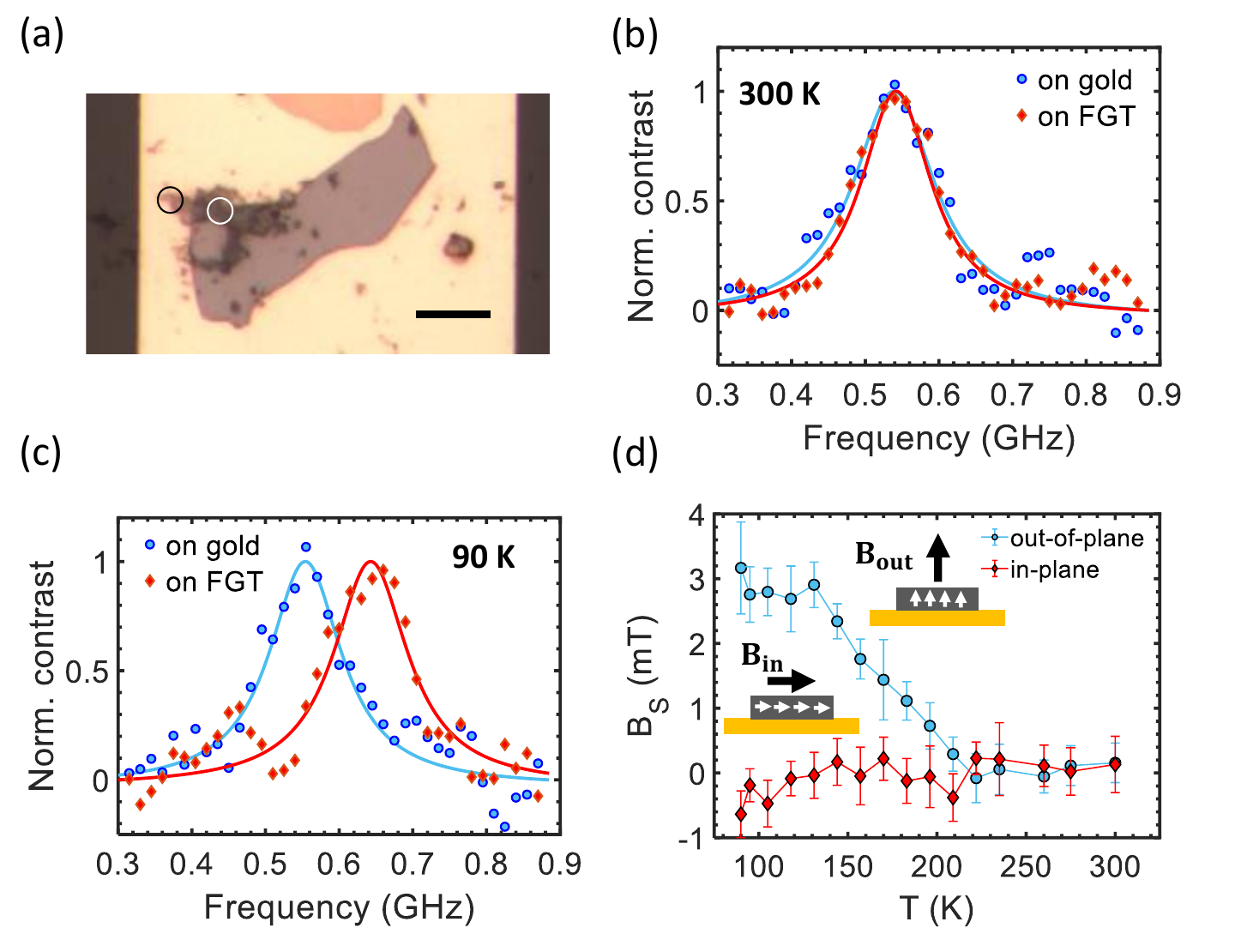}
	\caption{ \textbf{Characterization of  the stray magnetic field near a 2D magnet.} (a) An optical image of a BNNT cluster with spin defects on a Fe$_3$GeTe$_2$ (FGT) flake. A part of the BNNT cluster locates on the gold stripline for characterizing the background magnetic field. The white and black circles indicate the two spots for ODMR measurements. Scale bar: 10 $\mu$m. (b)-(c) Two examples of ODMR spectra taken with the BNNT spin defects on the gold (blue) and on the FGT (red) at (b) 300 K and (c) 90 K. The magnetic field is applied out-of-plane for (b) and (c). (d) The stray magnetic field generated by the FGT flake at different temperatures ranging from 90 K to 300 K. The external magnetic field is applied in-plane (red) or out-of-plane (blue).  } \label{fig4:2Dmagnets}
\end{figure*}

To demonstrate this capability, we perform a series of CW ODMR measurements, maintaining constant MW and laser powers while adjusting the magnetic field's orientation around two perpendicular axes (Figure \ref{fig3:magnetic}(a)-(b)). Regardless of the rotation angle, the spin resonance frequency remains constant, with minimal variance in ODMR contrast (Supplemental Fig. S7).   The minor contrast reduction seen in Figure \ref{fig3:magnetic}(b) arises from a decrease in the MW magnetic field component perpendicular to the static external magnetic field $B_{ext}$.  To  investigate the spin coherence properties of BNNT spin defects, we perform pulsed ODMR measurements to determine their spin relaxation times $T_1$. The pulsed ODMR sequence consists of an initial laser pulse for initializing the spin state, a MW pulse to manipulate the spin state, and a final laser pulse to readout the state. Using a resonant MW drive of varying durations as shown in Figure \ref{fig3:magnetic}(c), we observe a Rabi oscillation with a characteristic frequency $f_{R}$. After determining the Rabi frequency, spin relaxation measurements are performed for different magnetic field orientations. A measurement protocol is shown in the top panel in Figure  \ref{fig3:magnetic}(d).  The exponential decay time of PL gives the spin relaxation time $T_1$ of around 14 $\mu$s, which is nearly independent of the external magnetic field's orientation.

{\bf Sensing anisotropic magnetization}

To showcase the potential of omnidirectional magnetic field sensing, we employed BNNT spin defects to analyze the anisotropic magnetization of a representative 2D magnet.  Specifically, we use BNNT spin defects to directly measure the stray magnetic field $B_s$ generated by a 2D magnet with varied magnetization directions. Note it would be challenging to measure the stray magnetic field in orthogonal directions with a diamond NV center as it is insensitive to a field perpendicular to the NV axis. We found that only a few percents of bright emitters in commercial BNNTs are spin defects. To improve the probability of getting a spin defect in a BNNT, we treat  BNNTs with carbon ion implantation followed by thermal annealing for 2 hours. About 30$\%$ of bright emitters  in treated BNNTs show ODMR signals. Thus these BNNT spin defects are expected to be related to carbon atoms \cite{mendelson2021identifying}.

\begin{figure*}[htp]
	\centering
	\includegraphics[width=0.9\textwidth]{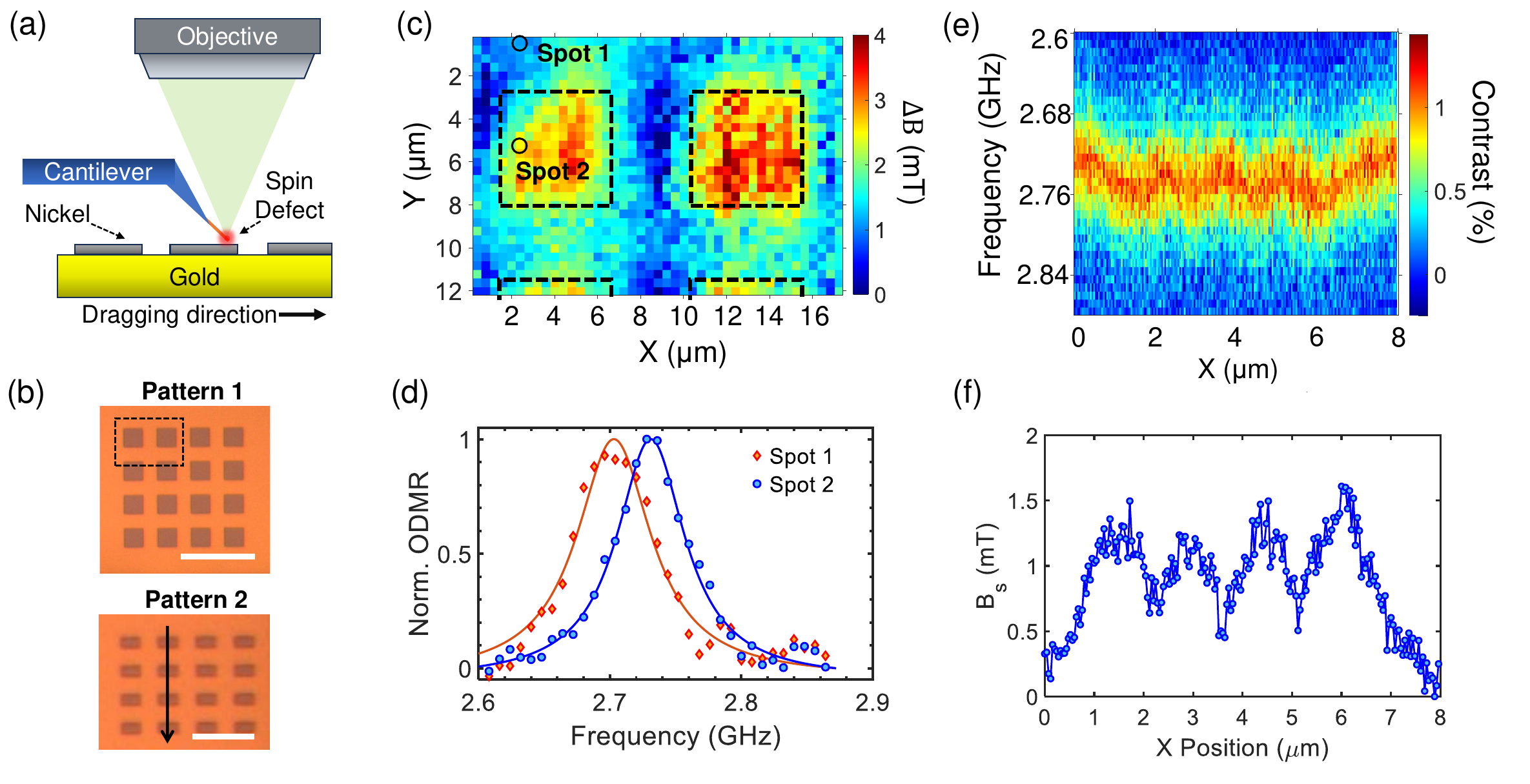}
	\caption{ \textbf{Scanning probe magnetometry with BNNT spin defects.} (a) A illustration of the setup for the scanning probe ODMR measurement. (b) The optical images of the Nickel pattern to be characterized. The scale bars are 15 $\mu$m in the top image and 10 $\mu$m (c) The stray magnetic field distribution measured by the scanning BNNT probe. The pixel size is 500 nm. (d) Two ODMR spectra measured away from (Spot 1) and on (Spot 2) the nickel patch. (e) The ODMR spectra as a function of position in a 1D scan along X axis (as depicted in the bottom panel of (b)). The step size is 40 nm. (f) The stray magnetic field measured based on the fitted resonance frequency from the data in (e).} \label{fig5:Scanning}
\end{figure*}

We start with an exfoliated Fe$_3$GeTe$_2$ (FGT) flake on a gold MW stripline \cite{tan2018hard,fei2018two}.  After transferring BNNTs with spin defects onto the FGT flake and the gold stripline (Figure \ref{fig4:2Dmagnets}(a)), the assembled device is mounted in a cryostat for temperature-varied ODMR measurements (Supplemental Fig. S9). The local magnetic field, $B_{loc}$, is extracted from the ODMR resonance frequency $f_{ODMR}$ through the Zeeman effect. Extracting the external field's contribution, $B_{loc,Au}$, as gauged by the ODMR on the gold stripline, gives the stray field from FGT: $B_s = B_{loc,FGT} - B_{loc,Au} =(f_{ODMR,FGT}-f_{ODMR,Au})/g\mu_B$. Figure \ref{fig4:2Dmagnets}(b)-(c) shows two ODMR spectra at 300 K and 90 K, in an external magnetic field $B_{ext}$=19.4 mT. At 300 K, the ODMR spectra gives nearly the same resonance frequencies for both FGT and gold, implying negligible stray magnetic field from FGT. In contrast, at 90 K, below a thick FGT flake's Curie temperature ($T_c \approx 220 K$) \cite{tan2018hard,fei2018two}, an  88.7 MHz ODMR frequency difference is observed. This corresponds to a stray field $B_ s \approx 3.2$ mT.  Figure \ref{fig4:2Dmagnets}(d) illustrates that as temperatures drop below about 220 K, the ODMR resonance frequency on FGT rises, while the frequency on gold remains stable, suggesting increased magnetization below the Curie temperature.  With an in-plane external magnetic field, the local stray field at the edge of FGT stays nearly unchanged within the measured temperature range. This shows the magnetization of FGT is highly anisotropic.

{\bf Scanning probe magnetometry}

BNNT spin defects, besides being placed directly on samples to sense stray fields, can also be attached to an AFM cantilever for scanning probe magnetometry, offering spatial resolution surpassing the optical diffraction limit. Notably, BNNTs are more than nanofabricated diamond tips with NV centers \cite{thiel2019probing,song2021direct}.  Here, we demonstrate  scanning probe magnetometry with BNNT spin defects attached on a cantilever to profile the stray field distribution over a nickel pattern. Figure \ref{fig5:Scanning}(a) depicts a schematic of the experimental setup. A BNNT with spin defects is attached to an AFM  cantilever and sticks out by a few micrometers. The scanning probe is brought close to the surface of a nickel pattern on a MW waveguide, maintaining a sub-micrometer gap.  For simplicity, our current setup does not have feedback stabilization found in commercial AFM systems. The nickel pattern on the waveguide is  scanned while the position of BNNT probe is fixed. ODMR spectra at each position determine the local magnetic field, with stray fields derived by deducting the minimum magnetic field in the scanned area.  

Figure \ref{fig5:Scanning}(b) shows two nickel patterns used in our experiment.
 For the $5 \mu m \times 5 \mu m$ pattern (Pattern 1) in Figure \ref{fig5:Scanning}(c), our 2D scan reveals a stray field map closely aligned with our simulation (Supplemental Figure S10). Two examples of ODMR spectra are presented in \ref{fig5:Scanning}(d), showing a clear shift in the spin resonance frequency when the probe is on top of the nickel patch. We further investigate the field distribution along a line across a smaller nickel pattern with a 600 nm width for each patch. The ODMR spectra as a function of the position with a step size of 40 nm is shown in Figure \ref{fig5:Scanning}(e). A clear shift in the ODMR resonance peak is resolved when the probe goes across the pattern. The stray field distribution obtained from the resonance frequencies matches well with the simulation (Figure \ref{fig5:Scanning}(f) and Supplemental Figure S10).

{\bf Conclusions}

In summary, we have observed single optically addressable spin defects in BNNTs with a suggested spin $S=1/2$ ground state. Unlike diamond NV centers, these BNNT defects enable magnetic field sensing in any orientation with consistent sensitivity. We use the BNNT spin defects for sensing stray magnetic fields in a wide range of temperatures. The inherent 1D structure of a BNNT with spin defects is cost-effective and optimal for scanning probes compared to nanofabricated diamond tips with NV centers or superconducting quantum interference devices \cite{vasyukov2013scanning}. Additionally, the absence of ZFS ensures the ODMR resonance frequency remains unaffected by temperature or strain, leading to magnetic field sensing that is resilient against environmental changes.

The spatial resolution of scanning probe techniques is primarily determined by the distance between the probe and the sample, which is a few hundred nanometers in this study. By implementing feedback stabilization and reducing this separation, we can significantly improve the resolution to the nanometer scale. Notably, when attaching a BNNT to an AFM cantilever, it remains stable even upon direct contact with the target (Supplemental Fig. S4), thanks to BNNT's resilience. This contrasts starkly with diamond NV scanning probes. Consequently, a contact mode could be feasible with BNNT spin defects, allowing for direct sample characterization without damage. A BNNT with a spin defect can image both the topological structure \cite{wilson2009carbon} and magnetic properties of a sample. Due to its small diameter, a BNNT can also be inserted into a living cell \cite{singhal2011multifunctional} to perform in situ quantum sensing. These unique features of BNNT spin defects open up innovative possibilities for atomic scale quantum sensing in diverse fields like condensed matter physics and biology.

\section*{Methods}

\textbf{ODMR experiment.} ODMR measurements were carried out using a home-built confocal microscope system. A 532-nm laser was sent through a 550 nm dichroic mirror and focused on the sample using a high numerical aperture (NA=0.9) objective lens with 100X magnification. An acousto optic modulator (ISOMET, M1205-T110L-1) was used as a fast optical switch. The PL was separated from the laser by the dichroic mirror and the residual laser light was blocked by two 550-nm long-pass filters. Afterward, the PL was coupled into an optical fiber, and guided to a single-photon counter (Excelitas, SPCM-AQRH).  The microwaves are generated by a Stanford Research Systems SG386 signal generator. The amplitude is modulated by two fast RF switches (Mini-Circuits ZASWA-2-50DRA+) and then amplified by an amplifier (Mini-circuits ZHL-10W-202s, or ZHL-16W-43-s+). For ODMR measurements, a pulse steamer (Swabian Instruments Pulse Streamer 8/2) sends pulses to modulate the RF switches , signal generator and AOM. A permanent magnet was mounted on a translation stage combined with a Goniometer behind the sample to apply a tunable external dc magnetic field. The temperature dependent measurements for characterizing stray magnetic fields generated from a Fe$_3$GeTe$_2$ flake were done in a cryostat using liquid nitrogen. The laser beam is guided into the cryostat chamber by using a flip mirror, enabling the ability to switch between the room-temperature setup and the low-temperature setups.

\textbf{Stray magnetic field sensing.} The external magnetic field was generated by a permanent magnet in this work. To change the direction of the external magnetic field applied to a Fe$_3$GeTe$_2$ flake in a cryostat, we first raise the temperature of the device to room temperature, change the direction of the magnet, and then cool down the system again. During this process, the Fe$_3$GeTe$_2$ flake is demagnetized at room temperature before it is magnetized again in a new external magnetic field. 
The quantization direction of a spin $S=1/2$  defect follows the direction of the total magnetic field.
When the magnitude of the external magnetic field, $B_{ext}$, is much larger than the magnitude of the stray magnetic field from the sample, $B_s$,  the BNNT spin defect will be primarily sensitive to the component of the stray magnetic field in parallel to the external magnetic field: $B_{s,\parallel}$. 
This is because $\sqrt{(B_{ext}+B_{s,\parallel})^2+(B_{s,\perp})^2}-B_{ext} \approx B_{s,\parallel}$ when $|B_{ext}+B_{s,\parallel}|\gg|B_{s,\perp}|$, where $|B_{s,\perp}|$ is the component of the stray magnetic field perpendicular to the direction of the external magnetic field. Thus by changing the direction of the external magnetic field, we can measure the stray magnetic field in different directions, following the orientation of the external magnetic field.

\textbf{Scanning probe magnetometry.} 
The scanning probe measurements were done in the same room-temperature setup but with an additional translation stage for mounting the cantilever with BNNT. Two piezo controllers (MDT693B) were used to control the positions of the cantilever and the sample individually. To perform scanning-probe ODMR measurements, the tip of the BNNT probe is first brought to a distance of 10 $\mu$m using the knobs on the stage manually. Further distance control is achieved by using the piezoelectric positioners, which allow us to make the probe-to-sample distance within a few hundred nanometers. 

\textbf{DC magnetic field sensitivity.} 
The DC magnetic field sensitivity for a single BNNT spin defect can be calculated for Lorentzian shape of the CW ODMR spectra as \cite{cai2023spin}
\begin{equation}
	\eta_{DC}\approx\frac{4}{3\sqrt{3}}\frac{h}{g\mu_B}\frac{\Delta\nu}{C\sqrt{I}},
\end{equation}
where $\Delta \nu$, $C$, $I$ are the ODMR linewidth, ODMR contrast and PL count rate, respectively. h is the Plank's constant, g $\approx$2 is the Land\'e factor, and $\mu_B$ is the Bohr magneton. Since the CW ODMR contrast and linewidth are nearly independent of the external magnetic field orientation, the sensitivity is mainly affected by the applied microwave power which affects the ODMR contrast $C$ and the linewidth  $\Delta \nu$. For a single BNNT spin defect on a gold stripline waveguide, the typical PL count rate is $\sim$ 250 kcts/s with a single-mode optical fiber (3 $\mu$m core size) to collect fluorescence
light. By replacing the single-mode fiber with a graded-index multimode fiber (Thorlabs Inc.) with a core size of 62.5 $\mu$m, the brightness is improved to $\sim$ 1.5 Mcts/s, with which we obtain the DC magnetic field sensitivity as shown in Supplemental Figure S8. In this work, the best DC magnetic field sensitivity $\eta_{DC}$ using a single BNNT spin defect is about 90 $\mu {\rm T}/\sqrt{\rm Hz}$. The brightness of our BNNT single spin defects is comparable to that of the ultrabright hBN single spin defects reported recently \cite{guo2023coherent}.


\section* {Acknowledgments}
We thank  Yong Chen and Igor Aharonovich for fruitful discussions. This work was supported in part by the Gordon and Betty Moore Foundation grant 10.37807/gbmf12259, and the National Science Foundation grant PHY-2110591.

%\section* {Author contributions}
%T. L., X.G. and S.V. conceived and designed the project. X.G., S.V., S.D. and Y.J. built the setup. X.G., S.V., S.D.  performed the experiment. X.G., S.V., and T.L. analyzed the data. S.D., P.J., K.S. and S.Z. prepared the samples. T.L. supervised the project. X.G., S.V. and T.L. wrote the manuscript with inputs from all authors. 


%\vspace{1cm}
%\bibliographystyle{unsrtnat}
%\bibliography{hBNquantumsensing}


\end{document}